\documentstyle[11pt,a4,epsf]{article} 

\def\be{\begin{equation}}
\def\ee{\end{equation}}
\def\bea{\begin{eqnarray}}
\def\eea{\end{eqnarray}}

\begin{document}%\LARGE
\rightline{April 1, 2000} 
\begin{center} {\Huge 

Galaxy Distributions and Tsallis Statistical 
Mechanics\\
}\vskip2truecm
{\large 
A. Nakamichi} \\ \sl{
Gunma Astronomical Observatory, Takayama, Agatsuma, Gunma 377--0702, Japan}\\
{\large 
I. Joichi} \\ \sl{
School of Science and Engineering,
Teikyo University,
Toyosatodai 1--1, Utsunomiya 320--8551, 
Japan}\\
{\large 
O. Iguchi, and M. Morikawa} \\ \sl{
Department of Physics, Ochanomizu University, Tokyo 112--8610, Japan}\\
\end{center}
\vskip3truecm

\begin{abstract} \noindent
Large-scale astrophysical systems are non-extensive 
due to their long-range force of gravity.
Here we show an approach toward the statistical mechanics of 
such self-gravitating systems (SGS). 
This is a generalization of the standard statistical mechanics 
based on the new definition of entropy; 
Tsallis statistical mechanics.
Developing the composition of entropy and the generalized Euler
 relation, 
we investigate the galaxy distributions in count-in-cell method.
This is applied to the data of CfA II South redshift survey.  
\end{abstract}\vskip1cm

\section{Introduction}%Equation Section (Next)
Astrophysical systems in the Universe are characterized by the gravitation.
The structure formed through this long-range force is quite different
from those formed through the other short-range forces.  
If the system does not strongly depend on the initial conditions 
of the Universe, we can apply statistical mechanics for describing 
such self-gravitating systems (SGS).
However, we cannot directly apply the standard Boltzmann statistical 
mechanics for SGS since the long-range nature of gravity strongly
violates the extensive property of the system 
which is the premise of statistical mechanics.
Actually, the total energy increases much faster than 
the particle number $N$, the partition function $Z$ often becomes 
complex\cite{iguchi99}, reflecting the fact that there is 
no absolute stable state in SGS.  

In order to seek for workable statistical mechanics of SGS, 
we try an approach based on the new definition of entropy 
whose extensivity is violated from the beginning; 
Tsallis statistical mechanics.  

We formulate the count-in-cell method 
for the large scale galaxy distributions in this new statistical mechanics.
First we calculate the expression of the composite entropy and 
the generalized Euler relation in this new statistical mechanics.
These are applied to the data of CfA II South redshift survey.  
The parameter q becomes negative, 
which represents the instability of gravity.  

\section{Tsallis Statistical Mechanics}%Equation Section (Next)
The ordinary Boltzmann statistical mechanics is characterized 
by the entropy $S =  - \sum\limits_i^{} {p_i^{} \ln p_i^{} } $.
The distribution function 
$p_{N,E}  = \exp \left[ { - \left( {E - \mu N} \right)/T} \right]/\Xi $ 
maximizes this entropy with constraints of the probability conservation, 
the energy conservation, and the particle number conservation.
This statistical mechanics is originally aimed to describe 
the multi-fractal and chaos structures. 
It is characterized by the entropy of the form \cite{tsallis88a} 
$S_q \left[ p \right] = \left( {\sum\limits_i {p_i^q  - 1} } \right)/(1- q)$, 
where $q$ is a real parameter.
Tsallis distribution function is obtained 
so that it maximizes this entropy with the same constraints.
The solution has a power law tail:
\be
p_{N,E} = 
  \frac{1}{{\Xi _q }}\left\{ {1 - \frac{{1 - q}}{{\tilde T}}
  \left( {E - \bar E - \mu \left( {N- \bar N} \right)} 
  \right)} \right\}_{}^{\frac{1}{{1 - q}}},
\label{p-ne}
\ee
where  $\tilde T \equiv CT$, and $C = \sum\limits_i {p_i^q }$.
Partition function is defined as 
\be
\Xi _q^{} = 
  \sum\limits_{N,E} {\left\{ {1 - \frac{{1 - q}}{{\tilde T}}
  \left( {E - \bar E - \mu \left( {N - \bar N} \right)} \right)} 
  \right\}_{}^{\frac{1}{{1 - q}}} }.
\label{grand-z}
\ee
Note that $p_{N,E}$ reduces to the ordinary Boltzmann form for $q \to 1$.
For the consistent formulation, it is important to notice that 
the observable expectation value is calculated 
by the escort distribution\cite{tsallis88b}: 
$P_i  = p_i^q /C$, 
$\left\langle Q \right\rangle _q  = \sum\limits_i {P_i^{} Q_i }$.
The averaged quantities $\bar E,\bar N$ in Eqs.(\ref{p-ne}) 
and (\ref{grand-z}) should be understood in this sense.  

\section{Galaxy Distribution in Count-in-Cell Method}%Equation Section (Next)
There are many approaches to describe large scale structure of 
the Universe.
In this paper, we will concentrate on distribution of galaxies.  
One of the well-known method to describe the distribution of galaxies is 
the two-point correlation function. 
However, when we solve equations of motions 
for the two-point correlation function, 
we need three-point correlation function. 
Such higher order essentialness is called BBGKY chain. 
Since we have to cut the BBGKY chain, 
we have to apply some kind of approximations. 

On the other hand, count-in-cell method is often used to describe 
distribution of galaxies. 
In this method, we use an analytic formula for probability $f(N)$ of 
finding $N$ galaxies in a randomly positioned volume $V$. 
As we don't have to use any approximation in count-in-cell method, 
we can apply it even for clustered system.       
Saslaw and Hamilton\cite{saslaw84}, and S. Inagaki introduced 
the virial parameter 
b=(gravitational correlation energy)/(kinetic energy of random motion), 
which measures the deviation from the dynamical-equilibrium.  
Then they found that in thermal-equilibrium, 
their theoretical investigation of $f(N)$ is consistent with 
he N-body simulations or catalogues of observations. 
Strictly speaking, we should not apply thermal-equilibrium theory for 
expanding Universe. However their consistency let us further study 
thermal-equilibrium statistical description. 
In evolution of the Universe, dynamical-equilibrium $b=1$ 
must be realized before the Universe reach thermal-equilibrium. 
Therefore in this paper we consider the dynamical-equilibrium $b=1$ case. 
We believe that there exist adequate fitting parameter other than 
the virial parameter $b$. 
That is the reason why we consider non-extensive statistics 
which gives us parameter $q$.   
    
Supposing the galaxies distribute according to Tsallis statistics, 
we consider a system described by equilibrium thermodynamics. 
That is the grand canonical ensemble, characterized by the given 
temperature and the chemical potential. 

The probability to find no galaxy in the volume $V$ is 
\be
f\left( 0 \right) \equiv \sum\limits_E {P_{0,E}^{} } = 
  P_{0,0}^{}  = \frac{{\left( {p_{0,0}^{} } \right)_{}^q }}{C},
\ee
where
\be
p_{0,0}^{} = 
  \Xi _q^{ - 1} \left[ {1 + \frac{{1 - q}}{{\tilde T}}
  \left( {\bar E - \mu \bar N} \right)} \right]_{}^{\frac{1}{{1 - q}}}.
\ee
The partition function can be decomposed as 
\be
\Xi _q^{} = 
  \sum\limits_{N,E} {\left( {1 - \frac{{1 - q}}{{\tilde T}}
  \left( {E - \bar E - \mu \left( {N - \bar N} \right)} \right)} 
  \right)_{}^{1 + q/\left( {1 - q} \right)} }  
          = \left( {\Xi _q^{} } \right)_{}^q C
\ee
and therefore 
$\Xi _q^{}  = C_{}^{1/\left( {1 - q} \right)} = 
\left( {1 + \left( {1 - q} \right)S} \right)_{}^{1/\left( {1 - q} \right)}$.  
Thus we obtain 
\be
f\left( 0 \right) = 
  \left( {1 + \left( {1 - q} \right)S} \right)_{}^{\frac{{ - 1}}{{1 - q}}} 
  \left( {1 + \frac{{1 - q}}{{\tilde T}}
  \left( {\bar E - \mu \bar N} \right)} \right)_{}^{\frac{q}{{1 - q}}} .
\label{f0}
\ee

In order to reduce the above expression, 
we need to calculate the composite entropy 
and the Euler relation in Tsallis statistics.
When we compose two systems A and B, 
the distribution function is given by 
$p_{ij}^{} (A,B) = p_i^{} (A)p_j^{} (B)$ and 
the composed entropy is 
$S_{A + B}^{}  = S_A^{}  + S_B^{}  + \left( {1 - q} \right)S_A^{} S_B^{}.$ 
Sequentially using this composition 
$S_{N + 1}^{}  = s + S_N^{}  + \left( {1 - q} \right)sS_N^{}$, 
(where $s$ means entropy of 1 system), 
we obtain the total entropy of N identical systems as   	
\be
S_N^{} = 
  \frac{{\left( {1 + \left( {1 - q} \right)s} \right)_{}^N  - 1}}{{1 - q}}.
\label{SN}
\ee

Generalized Euler relation is given by the following arguments. 
The variables $E,V,N$ are the natural arguments of the entropy: 
$S_N^{}  = S\left( {E,V,N} \right)$.
Differenciating 
$S_{\alpha N}^{}  = S\left( {\alpha E,\alpha V,\alpha N} \right)$ 
with respect to $\alpha$ and setting $\alpha  \to 1$, 
we obtain 
\be
\left. {\frac{{\partial S_{\alpha N}^{} }}{{\partial \alpha }}} 
\right|_{\alpha  \to 1}^{}  = \left. 
{\frac{{\partial S\left( {\alpha E,\alpha V,\alpha N} 
\right)}}{{\partial \alpha }}} \right|_{\alpha  \to 1}^{} 
   = \frac{E}{T} + \frac{{pV}}{T} - \frac{{\mu N}}{T},
\label{euler}
\ee
where we used 
$\partial S/\partial V = p/T,\;\partial S/\partial E = 1/T,
\;\partial S/\partial N =  - \mu /T$, 
which are guaranteed by the Legendre structure of 
the Tsallis thermodynamics\cite{plastino97}.  
In the above, we can put any other values for $\alpha $ as well.
In general, the $\alpha $-dependence is inherited from $S$ to $T$ 
on the right hand side of Eq.(\ref{euler}).
Non-extensivity of $S$ necessarily accompanies the non-intensivity of 
the Legendre-conjugate variable $T$.   
Actually, the fact that the variables $E,V,N$ are extensive and 
$p,\mu$ are intensive in Eq.(\ref{euler}) renders 
the $\alpha$-dependence of the temperature: 
\be
T\left( \alpha  \right) = 
  T_1^{} \left( {1 + \left( {1 - q} \right)s} \right)_{}^{1 - \alpha N} 
\label{T-alpha}
\ee
where $T_1^{}$ is the temperature of one particle system.
Note that $\tilde T \equiv CT$ defined just after Eq.(\ref{p-ne}) is 
$\alpha $-independent!  
Probably this quantity $\tilde T$ should be related with 
the physical temperature as defined from 
the velocity dispersion of the system.
However at present, we do not have idea on the meaning of 
the quantity $T$\footnote{
This situation is similar to the argument on 
the physical probability distributions at the end of the section two.
Among $P_i^{}$ and $p_i^{}$, related with each other by the factor $C$, 
we have chosen the escort distribution $P_i^{}$ 
as the physical distribution.  }.  
This point will be further discussed in the last section.  

Then we obtain the Euler relation in non-extensive statistical mechanics
\be
\frac{{\left( {1 + \left( {1 - q} \right)S} \right)
\ln \left[ {1 + \left( {1 - q} \right)S} \right]}}{{1 - q}} = 
   \frac{{E + pV - \mu N}}{T}.
\ee
This temperature on the right hand side is 
$T\left( {\alpha  = 1}\right)$ in Eq.(\ref{T-alpha}), 
though the explicit form of the temperature does not appear 
in the final expression of $f\left( N \right)$.    
Using this Euler relation, we can now eliminate 
$\bar E - \mu \bar N$ in $f\left( 0 \right)$ and Eq.(\ref{f0}) reduces to
\be
f\left( 0 \right) = 
  \left( {1 + \left( {1 - q} \right)S} 
  \right)_{}^{\frac{{ - 1}}{{1 -q}}} 
  \left( {1 - \frac{{pV}}{{\tilde T}}\left( {1 - q} \right) + 
  \ln \left[ {\left( {1 + \left( {1 - q} \right)S} \right)} \right]} 
  \right)_{}^{\frac{q}{{1 - q}}} .
\ee
Since we consider the dynamical-equilibrium system, 
the system is fully virialized $b=1$, 
and therefore the pressure $p$ must be $0$, we obtain: 
\be
f\left( 0 \right) = 
  \left( {1 + \left( {1 - q} \right)S} 
  \right)_{}^{\frac{{ - 1}}{{1 - q}}} 
  \left( {1 + \ln \left[ {\left( {1 + \left( {1 - q} \right)S} \right)} 
  \right]} \right)_{}^{\frac{q}{{1 - q}}} .
\ee
Moreover using Eq.(\ref{SN}), we finally obtain
\be
f(0) = 
  \left\{ {1 + \left( {1 - q} \right)s} 
  \right\}_{}^{\frac{{ - N}}{{1 - q}}} 
  \left[ {1 + N\ln \left\{ {1 + \left( {1 - q} \right)s} \right\}} 
  \right]^{\frac{q}{{1 - q}}} .
\ee
Our parameters are $s$ (the unit of non-extensive entropy per galaxy) 
and $q$ (the Tsallis statistical parameter).   

Probability of finding $N$ galaxies in the volume V is 
generated from $f\left( 0 \right)$:\footnote{ 
Note that this expression is slightly different from that 
by Saslow and Hamilton \cite{saslaw84}.  }
\be
f\left( N \right) \equiv \frac{{( - n)^N }}{{N!}}\frac{{d^N }}{{dn^N }}f(0)
\ee
where $n$ is the galaxy number density\cite{white79}.
This is because the void probability $f\left( 0 \right)$ contains 
all the information of the whole correlation functions:
\be
f\left( 0 \right) = 
  1 - \int_{}^{} {P\left\{ {X_1^{} } \right\}}  + \int_{}^{} {\int_{}^{} 
  {P\left\{ {X_1^{} X_2^{} } \right\}} } - 
  \int_{}^{} {\int_{}^{} {\int_{}^{} {P\left\{ {X_1^{} X_2^{} X_3^{} } 
  \right\}} } }  \pm  \cdots ,
\ee
where $P\left\{ {X_1^{} X_2^{} } \right\}$ is 
the probability that there are galaxies in $dV_1^{}$ at $X_1^{}$ and 
$dV_2^{}$ at $X_2^{}$. 
The above expression guarantees that the probability is 
properly normalized:
$\sum\limits_{N = 0}^\infty  {f\left( N \right)}  = 1$.  

\section{Comparison with observations}%Equation Section (Next)
We use the data of CfA II South redshift observations 
which includes 4392 galaxies \cite{huchra98}.  
We have to reduce the data to the uniform sample. 
First we restrict the data to the galaxies 
whose absolute luminosity is brighter than the magnitude -19.1 and 
the distance $4,000\sim 8,000 [{{\rm km}/\sec}]$.
The distance is measured by the cosmic redshift.
We further exclude the edge of the observation region.  
We applied the K-correction for compensating the reddening.
Finally the data is reduced to 870 galaxies.
The number density is 
$n = 4.44 \times 10_{}^{-9} [ \left( {{\rm km}/\sec } \right)_{}^{-3} ]$.

\begin{figure}[htbp] \begin{center} \leavevmode
\epsfysize=5.0cm
\epsfbox{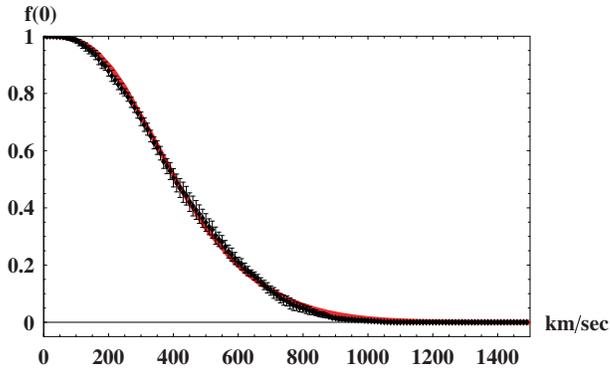}
\caption{
The void probability $f\left( 0 \right)$.
CfA II South Observations (dots with error bars) and 
our calculations (Solid lines) are plotted.  From this fit, 
we obtained the parameters q and s, 
which should be kept unchanged hereafter.}
 \label{fig-f0} \end{center} \end{figure}

\begin{figure}[htbp] \begin{center} \leavevmode
\epsfysize=18.0cm
\epsfbox{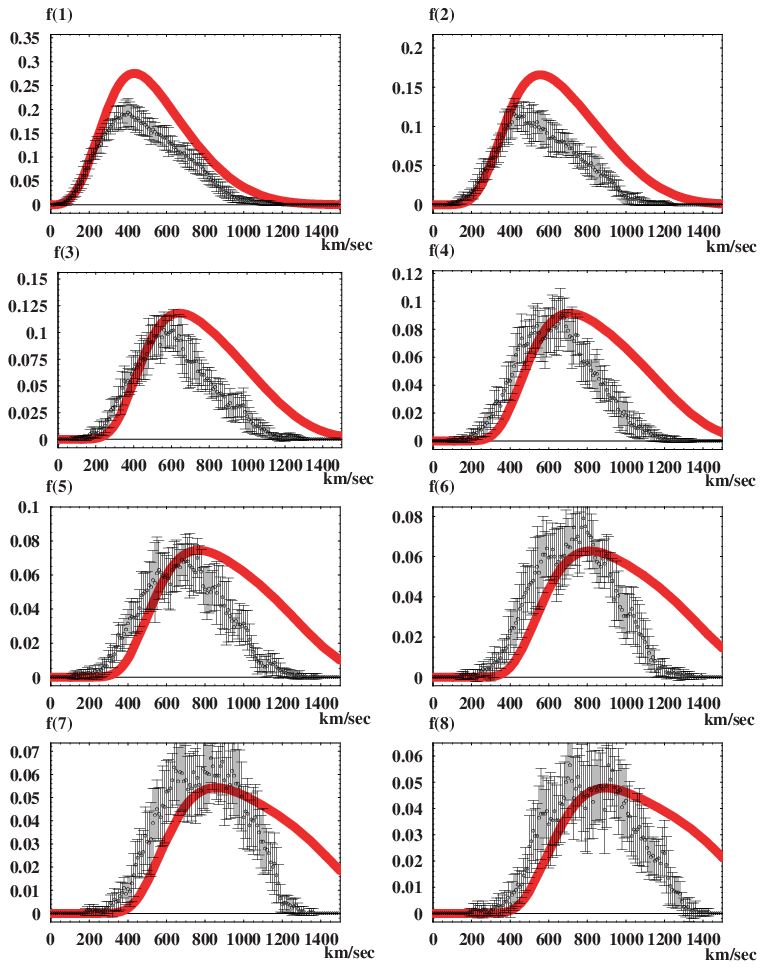}
\caption{
Probability functions $f\left( 1 \right)$ through $f\left( 8 \right)$, 
from left to right and top to bottom. 
CfA II South Observations (dots with error bars) and 
our calculations (Solid lines) are plotted.  
}
 \label{fig-f1-9} \end{center} \end{figure}

We first fit the void probability $f\left( 0 \right)$ 
by varying the parameters q and s. 
The best fit is realized by $q =  - 5.66847{\rm  }$ and 
$s = 0.164142{\rm  }$ (Fig.\ref{fig-f0}). 
We fix these values and do not change them hereafter in this paper.
With these parameters, 
general probability $f\left( N \right)$ is given. 
In Figs.\ref{fig-f0}-\ref{fig-f1-9}, 
we compared our calculations and the CfA data.      
%%%%%%%%%%%%%%%%%%%%%%%%%%%%%%%
%%%%%%%%%%%%%%%%%%%%%%%%%%%%%%%%
%%%%%%%%%%%%%%%%%%%%%%%%%%%%%%%
%%%%%%%%%%%%%%%%%%%%%%%%%%%%%%%%
We have checked the normalization of probabilities 
$\sum\limits_{N = 0}^\infty  {f\left( N \right)} = 1$ is realized.  

The negative value of $q$ we obtained may not be so surprising.
In reference \cite{costa97}, the value of $q$ appears to decrease 
monotonically from 1 to $ - \infty$ for one-dimensional logistic maps. 
Their model reveals unstable onset-to-chaos attractor.
For negative $q$, the entropy functional loses its convexity and 
the distribution becomes unstable. 
Thus the intrinsic instability of SGS is faithfully represented 
in this formalism and this fit.  

\section{Conclusions and Discussions}%Equation Section (Next)
We have constructed the non-extensive statistical mechanics 
based on the non-extensive entropy.  
Especially calculating the entropy of composite systems and 
deriving the generalized Euler relation in thermodynamics, 
we could evaluate the void probability function 
$f\left( 0 \right)$ and the probability to find 
$N$ galaxies $f\left( N \right)$.
This result was applied to the CfAII South galaxy observations and 
we have obtained negative parameter $q$.
This is thought to be another representation of 
the intrinsic instability of SGS.
It will be also interesting to notice the fact that 
the multi-fractal scaling is observed in this CfAII South data 
within the scale-region from $500[{\rm km/sec}]$ to 
$3000[{\rm km/sec}]$\cite{kurokawa99}.
We would like to clarify possible connection between 
the non-extensive distributions and 
the multi-fractal nature in the context of gravity.  

On the way we derive $f\left( N \right)$, 
we encountered ``scale ($\alpha $) dependent temperature $T$''.
If we put the values we obtained $q =  - 5.66847{\rm  }$ and 
$s = 0.164142{\rm  }$ to Eq.(\ref{T-alpha}), 
we can explicitly plot the scale dependence of the temperature.  
It turns out to reduce with increasing scale and abruptly drops 
to zero at about $r \approx 600[{\rm km/sec}]$ or, 
assuming the cosmic expansion speed as $H = 72[{\rm km/sec/Mpc]}$, 
at about $8.3[{\rm Mpc}]$, 
which is almost the scale that 
the galaxy correlation function becomes unity.
On the other hand, it is apparent that 
the galaxies do have peculiar velocity of 
order $1000[{\rm km/sec}]$ at this scale.  
Therefore, at least, the quantity $T$ cannot be interpreted as 
the ordinary temperature as defined from the velocity dispersions.
One of our next task will be to elucidate the meaning of $T$.  

In relation with the astronomical velocity distributions, 
the authors\cite{lavagno98} claim that 
the velocity distribution of the {\it clusters of galaxies} 
can be well fitted by the Tsallis distribution with 
the parameter $q = 0.23_{ - 0.05}^{ + 0.07}$, 
which is apparently different from our negative value for 
{\it galaxy} distributions.
Further study on the velocity distributions in various scales 
(galaxies, clusters, super-clusters) would reveal 
the origin and evolution of the large-scale structure of Universe.
We would like to report these results in the near future.  

\section*{Acknowledgment}
I. J. would like to thank Prof. T. Yokobori and Prof. A. T. Yokobori, 
Jr for their hospitality.
All of us would like to thank S. Abe, M. Hotta and K. Sasaki 
for useful discussions and valuable comments.

\end{document}